\documentclass{article}
\usepackage{arabtex}
\usepackage{utf8}
\setcode{utf8}
\usepackage{arxiv}
\usepackage[utf8]{inputenc} % allow utf-8 input
\usepackage[T1]{fontenc}    % use 8-bit T1 fonts
\usepackage{hyperref}       % hyperlinks
\usepackage{url}            % simple URL typesetting
\usepackage{booktabs}       % professional-quality tables
\usepackage{amsfonts}       % blackboard math symbols
\usepackage{nicefrac}       % compact symbols for 1/2, etc.
\usepackage{microtype}      % microtypography
\usepackage{lipsum}		% Can be removed after putting your text content
\usepackage{graphicx}
\usepackage{natbib}
\usepackage{doi}

\title{Query Logs Analytics: A Systematic Literature Review }
\author{ \href{https://orcid.org/0000-0002-3794-844X}{\includegraphics[scale=0.06]{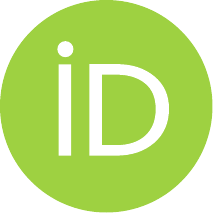}\hspace{1mm}Dihia LANASRI}\\
	ESI\\
	Algiers, Algeria\\	
	\texttt{ad\_lanasri@esi.dz} \\
}

\renewcommand{\shorttitle}{LOD Logs Analytics}

\hypersetup{
pdftitle={Query Logs Analytics Literature Review},
pdfsubject={NLP, ANALYTICS},
pdfauthor={D.LANASRI},
pdfkeywords={Linked Open Data, query-logs, Layered Architecture, end-to-end solution, Log Analytics},
}

\begin{document}
\maketitle

\begin{abstract}  %%OK Checked%%%
In the digital era, user interactions with various resources such as databases, data warehouses, websites, and knowledge graphs (KGs) are increasingly mediated through digital platforms. These interactions leave behind digital traces, systematically captured in the form of logs. Logs, when effectively exploited, provide high value across industry and academia, supporting critical services (e.g., recovery and security), user-centric applications (e.g., recommender systems), and quality-of-service improvements (e.g., performance optimization). Despite their importance, research on log usage remains fragmented across domains, and no comprehensive study currently consolidates existing efforts.

This paper presents a systematic survey of log usage, focusing on Database (DB), Data Warehouse (DW), Web, and KG logs. More than 300 publications were analyzed to address three central questions: (1) do different types of logs share common structural and functional characteristics? (2) are there standard pipelines for their usage? (3) which constraints and non-functional requirements (NFRs) guide their exploitation?. The survey reveals a limited number of end-to-end approaches, the absence of standardization across log usage pipelines, and the existence of shared structural elements among different types of logs.

By consolidating existing knowledge, identifying gaps, and highlighting opportunities, this survey provides researchers and practitioners with a comprehensive overview of log usage and sheds light on promising directions for future research, particularly regarding the exploitation and democratization of KG logs.
\end{abstract}

%%OK Checked%%%

\section{Introduction} %%OK Checked%%%
In the digital transformation era, data has become a critical asset shaping decision-making, automation, and innovation. From traditional enterprise systems to the Web, cloud platforms, and Knowledge Graphs (KGs), the exploitation of data has steadily evolved toward openness, scalability, and intelligence. Behind each interaction, whether you are performing an SQL query, browsing a website, or sending a request to a SPARQL endpoint, there remains a powerful hidden footprint: \textbf{logs}. These logs, initially conceived as low-level records of technical events, have progressively emerged as a strategic source of knowledge, enabling monitoring, optimization, personalization, and even predictive analytics.

The role of logs has undergone a remarkable transformation. Before the 1990s, information systems were mainly designed for internal use. Enterprise applications such as ERP, HR, and financial systems generated structured data stored in relational databases and exploited by known users within a closed organizational environment. Logs in this context were closed-world, reflecting controlled usage scenarios and trustworthy provenance. With the advent of the Web, IoT, and the Semantic Web, this paradigm radically shifted: applications became global, data became open, and millions of users worldwide started interacting with systems. This openness gave rise to open-world logs—search queries, clickstreams, API calls, and server events—that are abundant, heterogeneous, and harder to trust. Unlike closed logs, they introduce critical challenges in terms of quality, provenance, and security, making their exploitation both promising and risky.

Today, logs exist in nearly every layer of the digital ecosystem:
(i) Database logs capture user queries, transaction histories, and optimization hints; 
(ii) Data warehouse logs store records of analytical queries and workloads for large-scale reporting;
(iii) Web and server logs record navigation paths, clicks, and events, providing insights into behavior and performance;
(iv) System and network logs are essential for monitoring reliability, detecting anomalies, and ensuring security;
(v) Knowledge Graph (KG) logs trace SPARQL queries or interactions through QA/chatbot systems, reflecting the growing importance of semantic technologies.

Such diversity highlights the multifaceted nature of logs: they can be exploited for system-centric goals (availability, recovery, performance optimization), for user-centric services (recommendation, personalization, search enhancement), or for research-driven objectives (workload characterization, benchmarking, machine learning model training). This richness explains why logs have attracted attention since the early days of computing, but also why their study remains fragmented across domains such as databases, web mining, and AI.

A striking observation from the literature is the absence of a unified vision. While many works address logs in specific domains, there is still a lack of systematic surveys providing a global perspective on their exploitation pipelines—covering preprocessing, storage, curation, analysis, and their relation with Non-Functional Requirements (NFR) such as scalability, trust, and reproducibility. This fragmentation limits the democratization of logs, as most pipelines remain proprietary, domain-specific, or poorly documented.

This gap becomes even more evident when looking at KG logs. Knowledge Graphs have become a backbone technology in both academia and industry, powering semantic search engines, recommendation systems, and natural language interfaces. Initiatives like DBpedia, Wikidata, and enterprise KGs have fueled their adoption. Yet, the logs they generate remain an underexplored resource. Despite containing valuable traces of user intent, query complexity, and interaction patterns, KG logs have received limited attention in terms of systematic exploitation. Only a few works have explored query analysis \cite{}, user-centric services \cite{}, or benchmark creation \cite{}, leaving significant potential untapped.

To address this situation, this paper proposes a \textit{systematic survey} of log analytics, focusing on four major categories: Database (DB) logs, Data Warehouse (DW) logs, Web logs, and Knowledge Graph (KG) logs. Covering over 300 research works, our objectives are to:
\\- Provide a structured taxonomy of log exploitation approaches across different domains, identifying commonalities and divergences;
\\- Analyze methodologies and pipelines, including preprocessing, storage, and analytical techniques, as well as constraints and NFRs;
\\- Highlight the research gap in underexplored areas, particularly KG logs, and showcase their potential for future developments.

By bridging scattered contributions, contrasting closed-world and open-world perspectives, and drawing attention to the rising importance of KG logs, this survey aims to serve as both a reference point for researchers and a roadmap for practitioners. It emphasizes that logs should no longer be viewed as passive by-products of computation, but rather as first-class citizens in data-driven innovation and a cornerstone for trustworthy, scalable, and user-centric systems.

\section{Research Methodology} %%OK Checked%%%
\begin{figure*}[h] 
	\centering	 
	\includegraphics[width=17cm, height=6cm]{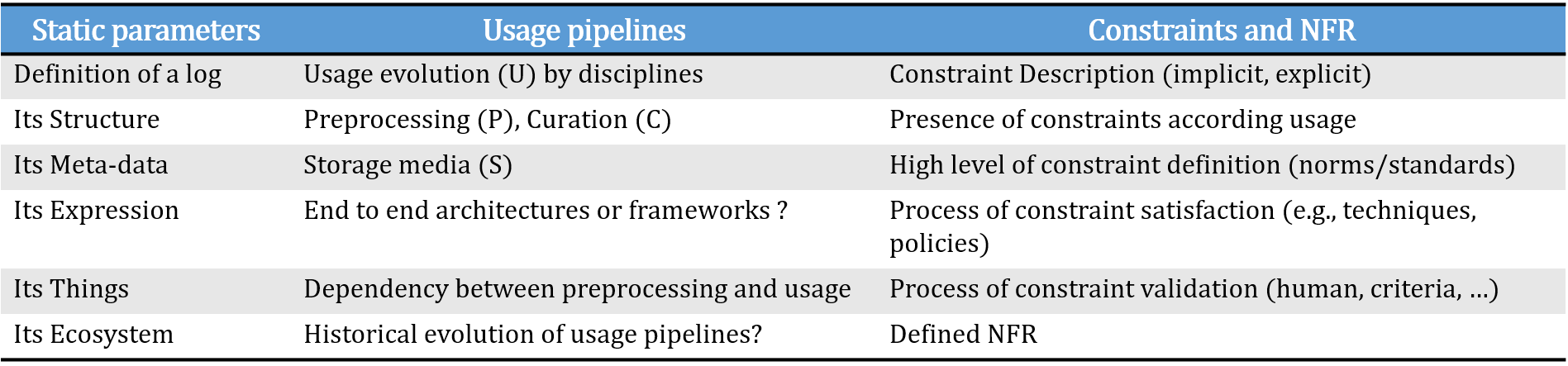}
	\caption{Research Questions}
	\label{questions}
\end{figure*}

To ensure a systematic and reproducible process, we adopted a structured methodology inspired by the principles of systematic literature reviews (SLR). The goal was to collect, filter, and analyze the most relevant contributions related to log analytics across different domains (databases, data warehouses, web, and knowledge graphs). The methodology followed five main steps, as detailed below:

\textbf{1. Define Research Questions (RQs).} The survey was guided by a set of well-defined research questions aimed at structuring the investigation and delimiting its scope. These questions served to clarify the objectives, identify the dimensions of analysis, and ensure that the selected works contribute to answering them.
The reserach questions are resumed in the figure \ref{questions}.

\textbf{2. Select Search Engines and Digital Libraries.}
The literature search was conducted using three widely recognized and complementary academic search engines:
\\- Google Scholar: for its broad coverage, including cross-disciplinary works and highly cited papers.
\\- DBLP: to ensure comprehensive coverage of computer science–oriented publications, including conference proceedings.
\\- Semantic Scholar: for its AI-assisted indexing capabilities, enabling the discovery of relevant works that might be missed otherwise.
This combination allowed us to balance breadth, relevance, and precision in retrieving the literature.

\textbf{3. Define Keywords and Search Queries.}
The identification of appropriate keywords was a crucial step to ensure comprehensive coverage of the relevant literature while keeping the scope consistent with our research questions. The process was conducted in many sub-phases:

- \textbf{\textit{Identification of Disciplines}}: We targeted four major domains where logs are generated and exploited:
(i) Databases (DB),
(ii) Data Warehouses (DW),
(iii) Web Systems, and
(iv) Knowledge Graphs (KGs).
These disciplines represent both mature and emerging fields where log analytics plays a significant role.

- \textbf{\textit{Identification of Relevant Keywords}}: Based on prior studies, preliminary readings, and domain-specific terminologies, we defined a set of core terms such as log analytics, query logs, workload analysis, log mining, and usage data.

- \textbf{\textit{Expertise and Conference Sessions}}: Keywords were also derived from domain expertise and by reviewing conference sessions and journal special issues focused on related themes. Top venues in databases, data management, and semantic technologies (e.g., VLDB, SIGMOD, WWW, ISWC, ESWC) provided valuable guidance on the most frequently used terminology.

- \textbf{\textit{Equivalent Terms and Synonyms}}: To maximize recall and capture variations in terminology, we also included equivalent expressions such as workloads, user traces, user history, past queries, and query workloads. These synonyms ensured that relevant works were not missed due to vocabulary differences between sub-communities.

Boolean operators (e.g., AND, OR) and combinations of domain-specific terms (e.g., “query logs” AND “knowledge graphs”, “log analytics” AND “data warehouses”) were employed to generate comprehensive queries across the selected search engines.

\textbf{4. Define Eligibility Criteria.}
To ensure both the quality and the relevance of the selected literature, a set of eligibility criteria was defined and systematically applied during the screening phase. These criteria covered publication type, venue ranking, publication date, language, keyword occurrence, and citation impact. The adopted criteria are as follows:

- \textbf{\textit{Type of Publication}}: Only peer-reviewed conference papers and journal articles were considered. Other sources such as book chapters, dissertations, and non-peer-reviewed reports were excluded.

- \textbf{\textit{Venue Quality}}:
* Conferences: restricted to CORE-ranked venues (A*, A, B), covering leading events in databases, knowledge management, and web technologies.

* Journals: restricted to those ranked Q1 or Q2 in Scimago Journal Rankings, ensuring high-quality and impactful publications.

- \textbf{\textit{Publication Period}}: Studies published between 1990 and 2022 were considered. This range was chosen to capture early foundational work on log analysis (1990s) as well as recent advances in machine learning and knowledge graph–oriented analytics.

- \textbf{\textit{Language}}: Only papers written in English were included to maintain consistency and accessibility.

- \textbf{\textit{Keyword Occurrence}}: Selected papers were required to contain at least one of the predefined keywords or equivalent terms in the title, abstract, or body text.

- \textbf{\textit{Citation Threshold}}: To filter impactful and recognized works, we applied a citation-based threshold:

* For older papers (published more than 10 years ago), a minimum of 20 citations was required.

* For recent papers (published within the last 10 years), a minimum of 5 citations was required.

These criteria ensured the inclusion of both seminal contributions that shaped the field and recent high-quality works reflecting the latest research trends.

\textbf{5. Data Collection and Analysis}
After applying the defined eligibility criteria, a final corpus of \textbf{303 papers} was selected for detailed analysis. The distribution of these papers is as follows:

\textbf{\textit{By type of publication:}} Conference papers: 173; Journal articles: 130

\textbf{\textit{
By domain:}} Databases (DB): 86; Data Warehouses (DW): 45;  Web: 133; Knowledge Graphs (KG): 39

\begin{figure*}[h] 
	\centering	 
	\includegraphics[width=17cm, height=8cm]{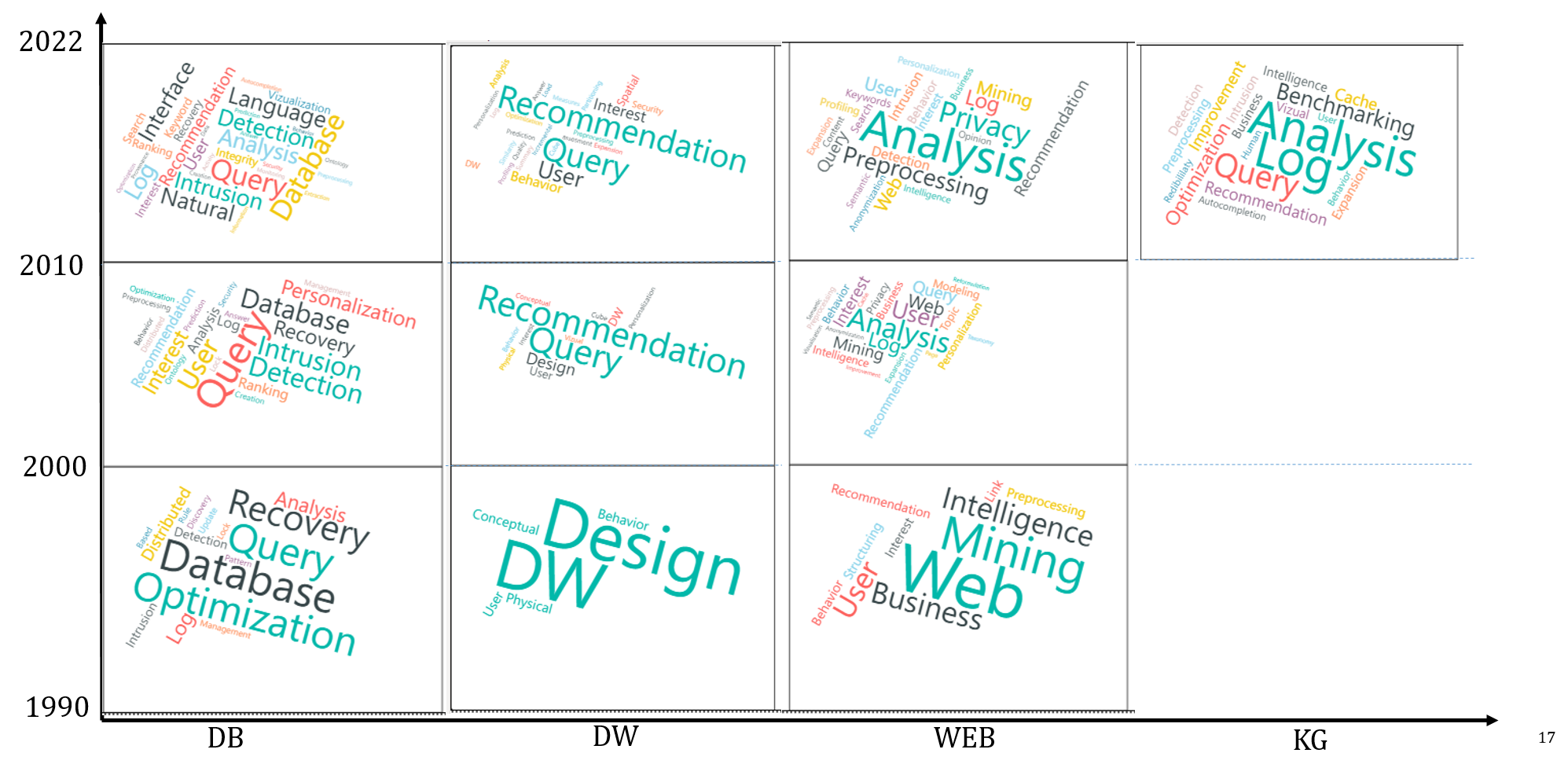}
	\caption{Historical Evolution of Query Logs usage}
	\label{history}
\end{figure*}

This distribution reflects the historical evolution and research focus of the community. While databases and data warehouses received early attention, the web domain dominates in terms of the number of studies, highlighting the explosion of web usage logs with the rise of Web 2.0 and large-scale online platforms. Interestingly, knowledge graph logs (39 papers) remain relatively underexplored despite the increasing adoption of KGs in industry and academia, confirming the existence of a research gap that this survey aims to address. 
The historical evolution od Query logs usage is resumed in the Figure \ref{history}.

The retained studies were systematically analyzed according to a predefined set of criteria:

- Application purposes (system monitoring, performance optimization, user modeling, etc.),

- Preprocessing and storage strategies,

- Consideration of non-functional requirements (NFRs) such as scalability, trust, reproducibility,

The results of this process were synthesized into a comprehensive taxonomy and thematic discussion, highlighting both mature areas and underexplored research gaps—particularly concerning Knowledge Graph logs.

\section {Results of Literature Review}
Query logs have been extensively studied across different domains, particularly in the contexts of databases, data warehouses, and the Web. Recently, Knowledge Graph (KG) logs have emerged as a promising but relatively underexplored source of information. In this section, we provide an overview of the main efforts dedicated to query-log exploitation. We then devote special attention to KG logs, which constitute the core of our study.  

The objective of this analysis is threefold:  
\begin{enumerate}
    \item Identify the essential components that constitute query logs.  
    \item Examine the exploitation pipelines adopted in the literature, generally involving the steps of \textit{acquisition}, \textit{preparation}, \textit{curation}, \textit{storage}, and \textit{usage}.  
    \item Highlight the major constraints that must be considered for developing query-log driven solutions, with a particular emphasis on issues of \textit{trust}, \textit{privacy}, and \textit{security}.  
\end{enumerate}

\subsection{Query-Logs in the Worlds of Data Repositories and the Web}
This section presents the main studies proposing to manage query-logs in the worlds of data repositories (transactional and analytical databases) and the Web. 

\subsubsection{Essential Components of Query-logs}
\label{component}
The components of query-logs are relevant objects that should be identified and analysed because they impact the different pipelines of their exploitation. Each query-log contains many records which represent the interaction of users with a data source. One important element of query-logs concerns their format which slightly changes from one data source type to another. 

In \textbf{\textit{transactional logs}}, each record represents the SQL query text associated with these main metadata: 
\textit{$<$execution datetime, user name, database name, table name, database server$>$.}

In \textbf{\textit{decisional databases}}, two main types of query-logs are distinguished based on the type of the query language used (SQL or MDX): 
\begin{itemize}
    \item In SQL query-logs, we find the analytical SQL query text associated with the following metadata: \textit{$<$execution datetime, user name, data warehouse name, data warehouse server$>$}.
    \item In MDX (OLAP) query-logs, in addition to the above metadata, we find new metadata describing \textit{$<$facts, dimensions, attributes, OLAP database name, OLAP server$>$} and the MDX query text.
\end{itemize}

In the web, two main types of query-logs are available:  
\begin{itemize}
    \item Navigational logs which contain the links visited by users. 
    \item Web query-logs of search engines which contain the keywords used by users for their search.
\end{itemize}

Both logs are associated with these metadata \cite{baglioni2003preprocessing}: \textit{$<$IP address, user session, execution datetime$>$}, and navigational logs have additionally these metadata: \textit{$<$http status, response size, cookies, Referrer$>$}. 

\begin{figure*}[h] 
	\centering	 
	\includegraphics[width=17cm, height=8cm]{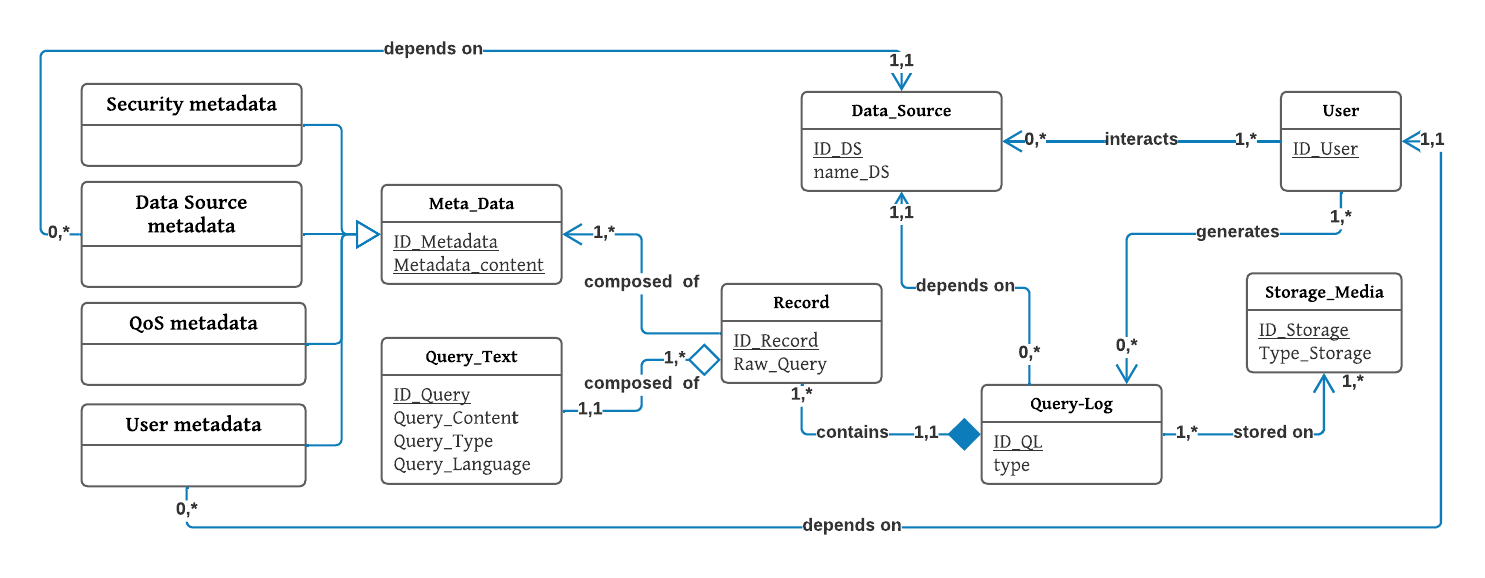}
	\caption{Query-Log Conceptual Model}
	\label{logsmodel}
\end{figure*}

To clarify the main components of a query-log in these worlds, we defined the query-log model proposed in Figure \ref{logsmodel}. As illustrated in the model, the interaction of users with a data source like a transactional database, OLAP database, etc. generates query-logs. Once generated, query-logs are stored on any storage media like files, databases, etc. to preserve them for any usage case. These query-logs contain many records and each record represents a raw query. The analysis of these different query-logs shows that they share two main components: 

\begin{enumerate}
    \item A query text belonging to a given type like textual in case of web search logs or structured in case of data repositories query-logs which are written in a given query language (e.g. SQL, MDX, etc). 
    \item Metadata that can be classified into four categories:
        \begin{enumerate}
            \item User metadata (e.g., user name)
            \item Data source metadata (e.g., database name, tables name)
            \item Security metadata (e.g., IP address, server name)
            \item QoS metadata (e.g., the execution datetime)
        \end{enumerate}
\end{enumerate}

User metadata, security metadata and QoS metadata are available in all query-logs, while the data source metadata are not provided in web query-logs. 

\subsubsection{Query-Logs Pipelines}
The literature on query logs is abundant, both in academic and industrial fields. Most works agree that using raw query logs directly is challenging due to their heterogeneous structure, noisy nature, and provenance issues. Consequently, logs are often processed through a multi-stage pipeline. Our analysis reveals a recurring set of four essential stages:  
\begin{itemize}
    \item \textbf{Acquisition:} collecting query from different sources and preparing them for the next step.
    \item \textbf{Curation:} cleaning, normalization, and transformation of raw logs to improve quality. Enrichment, integration and annotation of logs to facilitate interpretation.  
    \item \textbf{Storage:} persistence of curated logs in appropriate storage systems for later reuse.  
    \item \textbf{Usage:} application of logs for diverse tasks such as query optimization, workload modeling, system benchmarking, and recommendation.  
\end{itemize}

While this general pipeline is widely recognized, its implementation varies significantly depending on the target domain (DB, DW, Web). Furthermore, despite numerous proposals, no unified architecture has yet been established to standardize query-log analytics across domains.

In this part, we detail the query-logs pipeline covering usage, preparation, curation, and storage in the studied worlds of data repositories and the web. 

\begin{figure*}[h!]    
	\centering 	
	\hbox{\hspace{-7 em} \includegraphics[width=20cm,height=17cm]{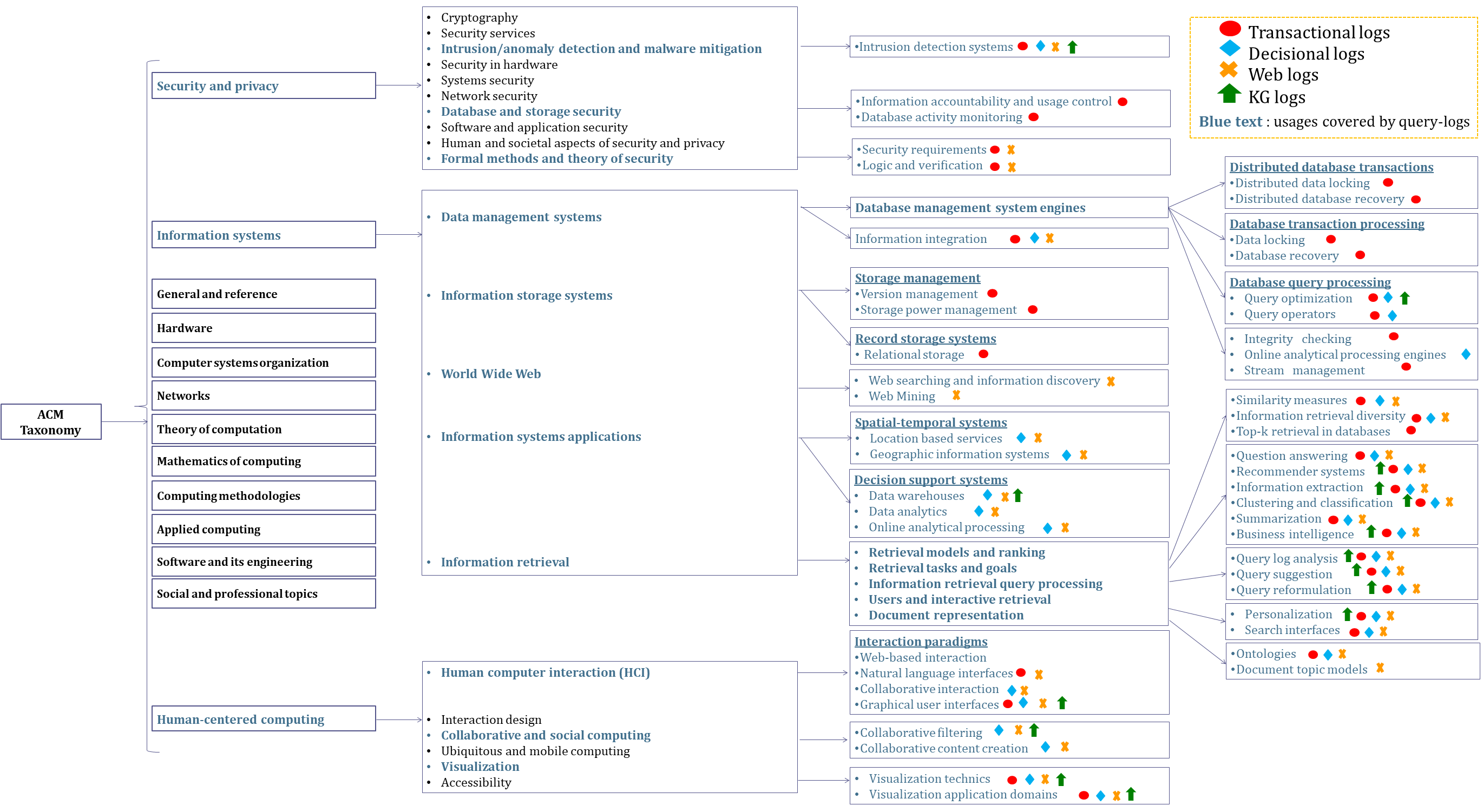}}
	\hspace{-0.2cm}
	\caption{Query-logs usage projected on ACM Taxonomy}
	\label{ACM}
\end{figure*}

\subsubsubsection{\textbf{3.1.2.1. Usage.}}\\
An exhaustive study of the important usages of query-logs is a complex task, as they are used in several domains and scientific disciplines. Consequently, we propose an intuitive vision that consists in projecting the studies related to query-logs on the ACM classification\footnote{\url{https://dl.acm.org/ccs}} as presented in Figure \ref{ACM} in order to enumerate their different usages. 

\textbf{Transactional databases:} Query-logs are traditionally associated with security and privacy management for intrusion/anomaly detection and malware mitigation \cite{low2002didafit}; database and storage security like database activity monitoring \cite{grushka2019simulating}, etc. They are also considered in data management systems for database transaction processing \cite{JimGrayBook} like distributed database recovery \cite{lomet1990recovery}, query optimization \cite{freytag1989basic}, and storage management \cite{schonig2019configuring,yang2011summary}.  

\textbf{Decisional databases:} Logs have been widely used, at the beginning, in physical design of decision support systems, where several optimization techniques such as indexes, materialized views \cite{bellatreche2000efficient}, and partitioning are selected based on these logs \cite{letrache2019olap}. They have contributed in conceptual design of data warehouses, where they were considered as functional requirements \cite{nair2007conceptual}.
Decisional query-logs have been considered for security and data management problems. They were strongly considered for Intrusion detection systems \cite{singh2013implementing}, information integration \cite{nair2007conceptual} and database query processing including classical and personalized OLAP queries \cite{bellatreche2005personalization}, etc.

\textbf{Web:} query-logs have been widely exploited to understand user preferences \cite{ramesh2017ontology} and to extract her profile \cite{ramesh2017ontology} using web mining techniques \cite{grace2011analysis}. Then, these extracted information are used to develop user-centric solutions satisfying user requirements grouped as Information retrieval solutions in ACM tree like recommender systems \cite{ramesh2017ontology} based on collaborative filtering \cite{suguna2013efficient} for query suggestion, web content \cite{baglioni2003preprocessing} and web caching personalization \cite{bonchi2001web}, query reformulation \cite{huang2009analyzing} etc. They are widely considered for information retrieval \cite{grace2011analysis}, enriching ontologies \cite{chuang2003enriching}, topic modeling \cite{jiang2013beyond}, etc. These logs were also exploited for security and privacy issues like intrusion detection systems \cite{doran2011web} and used in information integration solutions \cite{joshi2003using}, etc.

The immense advances made by the Web community have inspired database and data warehouse communities to consider analytical and transactional logs to develop user-centric solutions based either on user sessions and profiles \cite{giacometti2009recommending} or on collaborative filtering \cite{giacometti2009recommending} for creating recommender systems helping in SQL/MDX queries suggestion \cite{arzamasova2020sql,giacometti2009recommending} and recommending some parts of data cubes\cite{chanson2019profiling}; question answering systems \cite{baik2019bridging} based on natural language processing, content personalization \cite{sakka2021profile}, etc. 

Some Advanced analytics solutions were also proposed associated with visualization tools\cite{ahmed2015smart} and Graphical user interfaces \cite{francia2007visual} for deeper and graphical convivial analysis and collaborative interaction \cite{francia2020towards}. 

The arrival of Semantic Web (SW) with its different technologies used for advanced reasoning and semantic relations, has encouraged researchers to exploit these capabilities to improve the logs-based solutions. Using SW allowed deep understanding of users beliefs and identifying rich user profiles. It allows also enriching the discovered knowledge with external content \cite{ramesh2017ontology,ahmed2015smart} collected from domain ontologies and knowledge bases to return better and semantically enriched results from query-logs.

Figure \ref{ACM} points out the main ACM disciplines that exploit query-logs. Different colored symbols (see the legend) are used to project each query-log on the usages that it covers.

\subsubsubsection{\textbf{3.1.2.2. Preparation and Curation.}}\\

\textbf{Preparation and Curation}
\\ \label{curation}
Since query-logs are generated by users with different expertise levels, profiles and intentions, they suffer from several quality issues (e.g. wrong syntax, missing values). Therefore, they have to be prepared and cleansed to be used in different usage applications cited in Figure \ref{ACM}. Several approaches have been proposed in the literature covering our three worlds. Preparing query-log consists to extract the different components of a query-log (query and metadata) and propose many operations that help to improve its quality and verify its veracity.

In transactional logs, some studies proposed simple preprocessing pipelines to prepare these logs \cite{sobhan2002reorganization}, for solving quality issues by considering these logs in isolation way such as: defining the missing values, standardization of data types, and data conversion of the different extracted metadata. Other studies proposed complex pipelines \cite{low2002didafit} where multiple logs (with same or different structures) are aggregated, e.g. for merging many transactional logs extracted from many databases systems for strong intrusion detection. 

In data warehousing, curating query-logs consists to rewrite OLAP queries using a specific algebra \cite{aligon2012towards} or following a defined standard \cite{romero2011describing} for grouping sessions and queries needed for many usage cases like Recommender systems based on user and QoS metadata. Moreover, ontologies are used to identify the semantic of analytical queries \cite{ahmed2015smart}. These information are used for different usage cases like semantic recommendation \cite{ahmed2015smart}. 

Contrary to transactional and decisional query-logs which are generated by well-known users (internal and logged users), web query-logs may be generated by unknown users. Several studies proposed to manage the complexity of web query-logs \cite{baglioni2003preprocessing}, using different techniques dealing with: log preparation (e.g. most of the studies propose to extract the query text from the query-logs, and to separate metadata from the query-logs and organize them as fields \cite{lopes2015dynamic} which is called fields separation), log cleaning \cite{cooley1999data} (e.g. eliminate irrelevant items, delete bot/spider queries based on security metadata, deduplication of repeated queries), log transformation \cite{cooley1999data} (e.g. Normalization, data extension, data conversion, transaction/session/user identification and grouping based on user and QoS metadata, path completion, and new fields calculation \cite{lopes2015dynamic}, etc.). 

To summarize, the preparation of query-logs consists in separating the query text from the different metadata fields and parsing them to a human-readable format. While the curation solutions of the different query-logs can be classified into three main classes: (1) Cleaning which helps to eliminate all non valuable data (e.g. eliminate irrelevant items like photos and videos, eliminate bot/spider queries, deduplication); (2) Transformation aiming to enhance the representation of data (e.g. Normalization, data extension, data conversion, path completion, fields calculation, query correction); and (3) Merging/Integration which helps to group data according to a given pattern (e.g. transaction/ session/ user/ query grouping, log merging). The extracted metadata (mainly security, user and QoS metadata) play a crucial role to improve the quality of query-logs, they are considered in different curation solutions.

\subsubsubsection{\textbf{3.1.2.3. Storage.}}\\
Once prepared, and based on their quality and importance, query-logs are stored in: (1) traditional physical media \cite{yadav2012efficient} including files, databases, data warehouses; or recently in (2) semantic databases \cite{fernandez2016data} and knowledge graphs \cite{ekelhart2021slogert} to exploit their reasoning and inference capabilities.

With the rapid raise of query-logs volume mainly in web context, these storage repositories are associated with: Big Data infrastructures like HDFS \cite{priya2018analysis}, distributed NoSQL storage like MongoDB \cite{zheng2014big} and Big data technologies \cite{priya2018analysis} for parallel distributed processing in order to accelerate calculations and enhance the QoS of the proposed logs solutions.  

\subsubsection*{Synthesis}
Following a chronological analysis of the literature, query-logs have been predominantly considered for system-oriented tasks such as database management, security diagnosis, and query processing. Various types of metadata support these tasks, including security metadata for security diagnosis and QoS metadata for query processing and data management.

With the advent of the web, query-logs from data repositories and web sources began to be leveraged for user-centric solutions, primarily recommender systems and query personalization. In these approaches, the separation of query text and metadata fields is regarded as the main \textit{preparation} step, whereas user/session identification and grouping constitute the essential \textit{curation} tasks, enabling the construction of user profiles. In this context, user and QoS metadata are primarily exploited.

Several studies have proposed diverse pipelines for managing query-logs. Some works concentrate on a single curation task, such as session grouping, while others aim to enhance algorithms tailored to the intended application, for instance optimizing data mining algorithms for intrusion detection.

Despite these efforts, comprehensive end-to-end solutions encompassing query-log collection, preparation, curation, storage, and usage remain limited. Only a few works have proposed end-to-end frameworks or architectures \cite{francia2022cool, shinde2008new, zheng2014big}, which centralize the necessary tools within a single framework for effective analytics. A major limitation of these solutions is their specificity to a particular application (e.g., recommendation \cite{shinde2008new}, log analysis \cite{zheng2014big}), where the intermediate layers (preparation, curation, storage) are designed primarily to serve the final usage layer.

For example, in recommender systems or content personalization, certain curation tasks, notably session identification, are guided by the intended application, with user and QoS metadata playing a central role. For intrusion detection, data conversion focuses on security metadata to detect malicious queries. In business intelligence scenarios, normalization, data conversion, and log merging are the primary curation steps performed before storing the logs in a data warehouse for various decision-making processes.

\subsubsection{Constraints in Query-Log Exploitation}
Exploiting query-logs is challenging not only due to their structural complexity but also because of several critical constraints, including privacy, security, trust, and interpretability.

\paragraph{Privacy and Security} 
Privacy \cite{lauer2007building} and security concerns are frequently highlighted in the literature, given the sensitive nature of usage traces. Accessibility constraints also arise due to these concerns, limiting who can use and analyze query-logs. 

\paragraph{Trust} 
Trust is a complex and multidimensional concept encompassing quality \cite{ceolin2015linking}, provenance \cite{suriarachchi2016crossing}, privacy \cite{lauer2007building}, security \cite{artz2007survey}, credibility \cite{fogg2003users}, and reputation \cite{nepal2010behaviour}. Despite its critical importance, trust is rarely addressed explicitly in query-log research; most approaches assume logs are reliable without mechanisms to assess their quality or provenance. In transactional and analytical logs, trust is essential for accurate query optimization, auditing, and transaction processing. In contrast, in web logs, trust ensures reliable user modeling, recommendation, and search personalization \cite{suriarachchi2016crossing}. 

Trust is commonly enforced through verification, anonymization, and integrity-checking mechanisms, including access control and cryptographic techniques. In the broader literature, trust is generally defined as ''the subjective probability with which an agent expects that another agent or group of agents will perform a particular action on which its welfare depends'' \cite{gambetta2000can}. Both domain-specific \cite{ang2001trust} and generic conceptual models \cite{amaral2019towards} have been proposed to formalize trust and its components, which can be instantiated in various contexts \cite{lanasri2020trust}.

The manifestation of trust-related issues varies depending on the query-log environment:

\begin{itemize}
    \item \textbf{Transactional and Decisional Logs:} Generated in controlled, internal environments, these logs benefit from reliable provenance, controlled data sources, and authenticated users. Nevertheless, trust still requires attention. Studies have implicitly addressed trust through \textit{preparation and curation solutions} \cite{sobhan2002reorganization,romero2011describing} to improve quality, and \textit{intrusion or robot query detection} \cite{low2002didafit,singh2013implementing} to resolve provenance issues. Data repositories often include trust annotations, probabilistic databases \cite{cavallo1987theory}, and extended query languages (e.g., TrustQL \cite{ray2005vtrust}) to manage and verify trust.
    
    \item \textbf{Web Logs:} Web-generated query-logs pose greater trust challenges due to the unknown and potentially unreliable nature of contributors. Quality and provenance are major concerns, often addressed via curation and intrusion detection \cite{baglioni2003preprocessing,doran2011web}. Privacy is a central issue, as logs may include sensitive user data (IP addresses, session IDs, visited links), potentially compromising user identity or reputation \cite{poblete2007website}. Solutions include anonymization \cite{kumar2007anonymizing}, encryption \cite{jiang2010research}, and adherence to privacy policies \cite{cooper2008survey}. Additionally, frameworks like the Web of Trust \cite{caronni2000walking} have been proposed to enhance trust via authentication and security protocols.
\end{itemize}

\paragraph{Interpretability} 
The absence of graphical tools, reference ontologies, or standardized representations limits the accessibility of query-logs and makes their exploitation difficult for non-expert users.

\paragraph{Summary} 
Trust has been considered in all domains of interest, including both data repositories and web environments. While data repositories often include explicit trust annotations to ensure safe data usage, query-logs from these repositories generally lack explicit trust representation, addressing related issues such as quality and provenance only implicitly. Conversely, web query-logs often incorporate privacy and security mechanisms explicitly and enhance trust implicitly via quality and provenance management, but without formal trust annotations. These gaps underscore the need for systematic approaches to ensure observability, reproducibility, and safe exploitation of query-logs across contexts.

\subsection{Query-Logs in the world of KGs}
We consider our previous literature review of query-logs management in data repositories and web contexts as a prerequisite for identifying the main issues that should be treated in KG query-logs. We focus in this section on query-logs of KGs, by detailing their components, management pipelines and the notion of trust in the context of KGs. We conclude this literature review by projecting these efforts conducted in KG context w.r.t the issues identified in the previous section. 

\subsubsection{Essential Components of KG Query-Logs}
\begin{figure}[h!]
		\centering
		\includegraphics[scale=0.7]{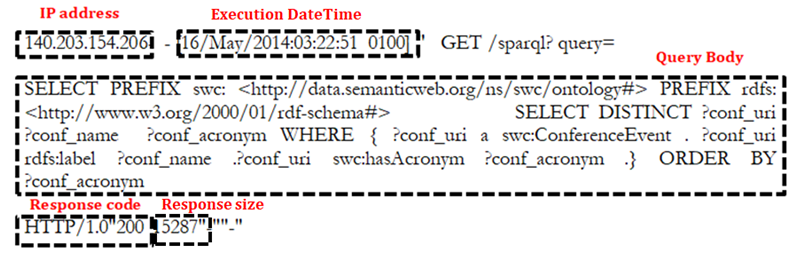}
		\caption{Structure of a KG query-log} 
		\label{log}
\end{figure}

KG query-logs record all users manipulations over KG data sources in form of many lines. KG query-logs structures depend on the SPARQL endpoint and the used services. Many efforts are conducted to define the structure of KG query-logs. Each line of KG query-log contains common parts with the previous query-logs (section \ref{component}) which are: the SPARQL query text executed by the user, associated with some metadata also classified into: i) user metadata: user ID (in some logs like DBpedia this information is provided) ii) Data Source metadata: KG name  iii) Security metadata: IP address (in some logs like Scholarly data this information is provided) iv) QoS metadata: like Execution DateTime, http response, and response size.  These metadata have a great added value during the curation and usage steps. They help in session identification, profile analysis, provenance analysis, etc.
To better illustrate these logs, Figure \ref{log} shows the structure of one line extracted from Scholarly Data KG query-log.

The conceptual model of KG query-logs is provided in Figure \ref{logsmodel}. 

\subsubsection{KG Query-Logs Pipelines}
\paragraph{Usages.} 
\label{Usages}
Over the last decade, several research efforts have addressed the exploitation of KG query-logs through statistical analysis, classification, and clustering techniques to understand their structure and content. Such analyses often support both graphical and semantic understanding of query-logs, using visualization tools and graphical interfaces such as DARQL \cite{bonifati2018darql} and SEMLEX \cite{mazumdar2011semlex} \cite{bonifati2020analytical}.  

KG query-logs have been leveraged for a variety of applications, which can be summarized as follows: 
\begin{enumerate}
\itemsep=0pt
    \item \textbf{Intrusion detection:} Identifying robotic and organic queries \cite{malyshev2018getting}.
    \item \textbf{Query-log analysis and information extraction:} Understanding the graphical representation and semantic content of query-logs \cite{bonifati2020analytical}.
    \item \textbf{Query optimization:} Improving source selection using data mining models that estimate the minimal set of sources needed to satisfy a query \cite{tian2011enhancing}.
    \item \textbf{Recommender systems:} Supporting users in query formulation through query suggestions based on collaborative filtering \cite{chen2014sparql}.
    \item \textbf{Query reformulation:} Using aggregated graph pattern ranking techniques to enhance queries \cite{rafes2018designing}.
    \item \textbf{Personalization and cache management:} Optimizing SPARQL endpoint caching and personalized query handling \cite{akhtar2020cache}.
    \item \textbf{Business intelligence:} Exploring multidimensional patterns from open KG query-logs for analytics, supported by interactive tools for data warehouse generation \cite{khouri2019loglinc, lanasri2019crumbs4cube}.
\end{enumerate}

These applications span several domains in the ACM Computing Classification System, as illustrated by the green up-arrows in Figure \ref{ACM}.

\paragraph{Preparation and Curation.} 
A review of existing works on SPARQL query-logs indicates that only minimal preparation and curation are generally performed to structure logs after extracting the main metadata \cite{mazumdar2011semlex}. These operations can be categorized into three main types:

\begin{enumerate}
\itemsep=0pt
    \item \textbf{Cleaning operations:} Deduplication of queries \cite{ell2011deriving,bonifati2018darql,bonifati2020analytical}, removal of incorrect or malformed queries based on quality metadata \cite{mazumdar2011semlex,ell2011deriving}, selection of relevant queries (e.g., \texttt{SELECT} queries) \cite{mazumdar2011semlex,ell2011deriving}, and extraction of RDF triples \cite{mazumdar2011semlex}.
    
    \item \textbf{Transformation operations:} Parsing KG query-logs \cite{mazumdar2011semlex}, identification of missing prefixes in incomplete queries \cite{ell2011deriving}, extraction of RDF triple features \cite{elbedweihy2011identifying,mazumdar2011semlex}, and correction of semantic and syntactic SPARQL errors \cite{almendros2017detecting}.
    
    \item \textbf{Merging and integration:} Unlike query-logs from data repositories and the Web, the merging and integration of KG query-logs have received little to no attention in the literature.
\end{enumerate}

These curated query-logs subsequently support various applications and usages, as discussed in Section \ref{Usages}.

\paragraph{Storage.} 
In the reviewed works, the storage of KG query-logs is rarely discussed in detail. In most cases, logs are simply stored in file-based structures. Although some preprocessing and curation operations are performed prior to storage, no study provides a comprehensive pipeline or dedicated architecture specifically designed for KG query-log analytics. Once stored, these curated logs are then used for the various applications and usage scenarios described in Section \ref{Usages}.

\subsubsection{Constraints in KG Query-Logs}
Trust has been widely studied in the context of KG datasets, but it has been only marginally considered for KG query-logs. In uncertain KG/KB datasets (e.g., YAGO \cite{suchanek2007yago}, Google Knowledge Vault \cite{dong2014knowledge}, NELL \cite{carlson2010toward}) each RDF triplet is annotated with a confidence or trust value, which informs users about the reliability of that triple \cite{djebri2019linking}. Approaches such as tRDF \cite{hartig2009querying} and languages like tSPARQL \cite{hartig2009querying} have been proposed to query these annotated data sources. Trust is closely related to quality \cite{ceolin2015linking}, and various measures—consistency, inter-linking, completeness—have been proposed to assess the quality and veracity of KG sources \cite{behkamal2015quality, zaveri2016quality}, including through deep learning techniques for KG completion \cite{wan2020adaptive}.

KG query-logs, however, exhibit issues that can affect their trustworthiness. Some works have addressed related problems, such as query-log preparation \cite{bonifati2020analytical}, semantic and syntactic correction of SPARQL queries \cite{almendros2017detecting} to improve quality, or bot query detection \cite{malyshev2018getting} to verify provenance. While these efforts tackle aspects related to trust, none have explicitly analyzed query-logs from a trust perspective. To date, no study has proposed annotating query-logs with trust values or modeling the concept of trust in this context.

\subsubsection*{Synthesis}
Compared to relational database or web search logs, Knowledge Graph (KG) query logs have received relatively little attention in the literature. Most existing studies focus on descriptive analyses of SPARQL queries, aiming to characterize their structure, frequency patterns, and graph-specific features. Only a few works have gone beyond analysis to propose user-centric services, such as recommender systems, query reformulation, or benchmarking.  

Despite these efforts, two critical gaps remain:  
\begin{enumerate}
    \item No dedicated architecture exists for the systematic exploitation and analysis of KG logs.  
    \item Trust—a central concern in open-world scenarios—is largely overlooked in the context of KG query logs.  
\end{enumerate}

This lack of comprehensive approaches contrasts with the growing importance of KGs in both academia and industry, where they support applications such as search engines, natural language interfaces, and recommendation systems. The scarcity of research on KG query logs motivates the need for a systematic investigation of their ecosystem and usage.

\section{Summary of the Survey Findings}
\begin{figure*}[h] 
	\centering	 
	\includegraphics[width=17cm, height=8cm]{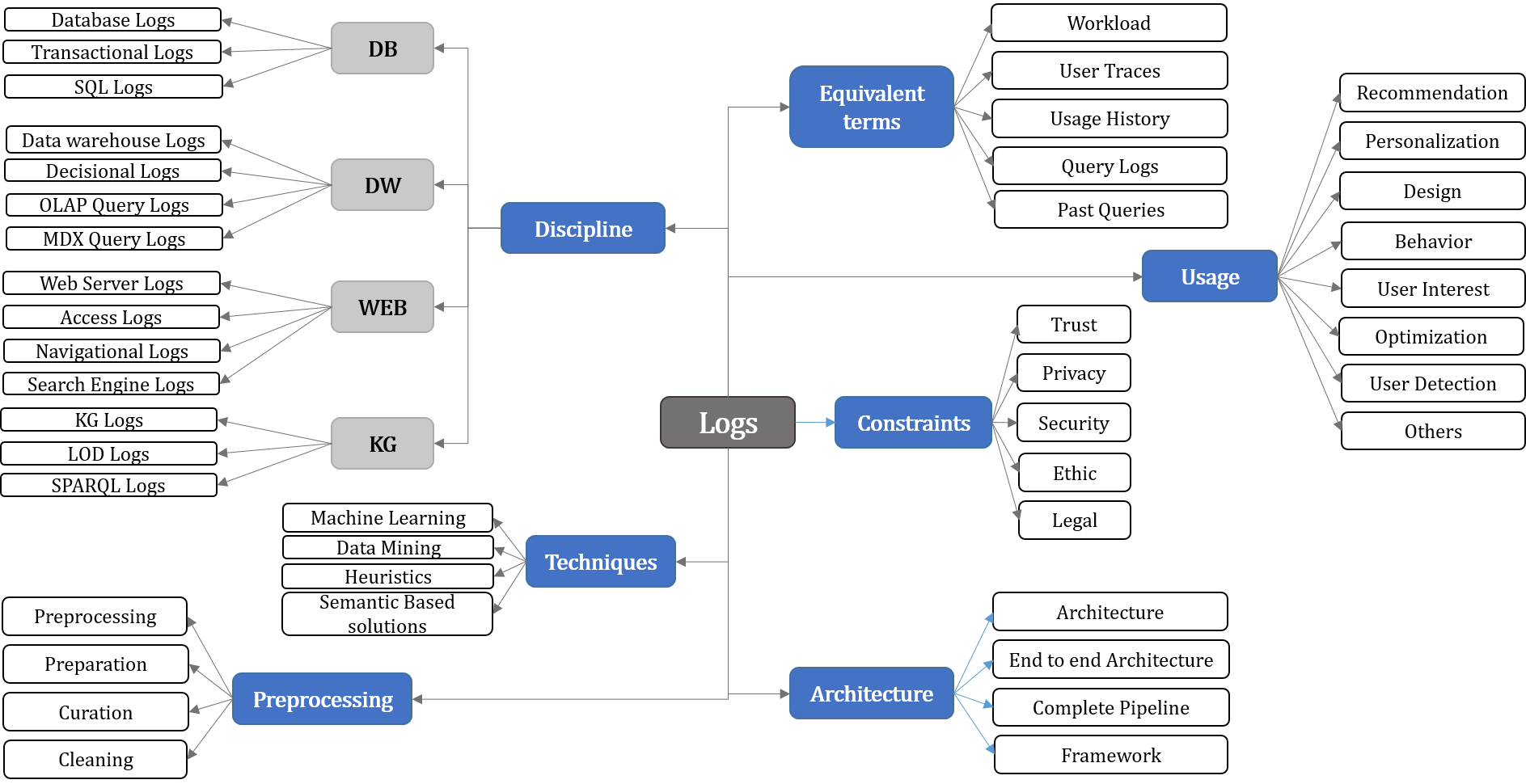}
	\caption{Survey MindMap}
	\label{mindmap}
\end{figure*}

Our survey of the literature on Knowledge Graph (KG) query logs reveals both progress and significant gaps, highlighting opportunities for future research and system design. The sum-up of this survey is given in the figure \ref{mindmap}. 
Several key insights emerge:

\begin{enumerate}
    \item \textbf{Raw KG query logs are not directly exploitable.} They must undergo a structured pipeline including preparation, curation, storage, and usage to become analytically meaningful.
    \item \textbf{Trust is largely neglected.} While privacy and security are sometimes addressed, trust, a critical concern in open-world and collaborative KG environments, is almost entirely overlooked in current analytics approaches.
    \item \textbf{Standardization is lacking.} The absence of shared tools, ontologies, meta-models, and visualization frameworks constrains reproducibility, observability, and explainability, limiting the practical impact of KG query-log analytics.
\end{enumerate}

Beyond these insights, our analysis identifies several recurring limitations:

\begin{itemize}
    \item Constraints (e.g., trust, privacy, quality) are rarely projected across the different layers of KG query-log pipelines. Most studies provide only global recommendations.  
    \item Operators and transformations within and between pipeline stages are insufficiently described, often failing to capture the specificities of SPARQL algebra and KG data structures.  
    \item Non-functional requirements, such as scalability, performance, and security, are better addressed in industrial deployments than in academic research prototypes.  
    \item Visualization tools, meta-models, and ontologies for managing constraints, ensuring data observability, and supporting explainability are largely absent.  
\end{itemize}

\paragraph{Towards an End-to-End Architecture for KG Query Logs.}  
These gaps motivate the design of a comprehensive, usage-agnostic architecture for KG query-log analytics, inspired by best practices from big data and log management systems. Such an architecture should cover:

\begin{enumerate}
    \item \textbf{Preparation:} Extraction of all relevant components from KG query logs, including SPARQL query elements, metadata, and user/session context.
    \item \textbf{Curation:} Profiling and detailed analysis followed by cleaning, transformation, enrichment, and deduplication operations, tailored to KG-specific characteristics.
    \item \textbf{Storage:} Scalable and structured storage solutions that support analytics, provenance tracking, and integration with downstream applications.
    \item \textbf{Usage:} Support for advanced analytics tasks, including query optimization, recommender systems, benchmarking, and pattern discovery, while embedding trust and reliability as first-class constraints.
\end{enumerate}

\paragraph{Trust as a Core Dimension.}  
To address the pervasive neglect of trust in KG query-log analytics, we advocate:

\begin{itemize}
    \item \textbf{Explicit modeling of trust:} Using UML or similar modeling languages to capture trust relationships, annotate SPARQL queries, and represent KG log ecosystems.
    \item \textbf{Layered propagation:} Trust annotations should propagate across all layers of the pipeline, enabling queries and logs to carry reliability metadata.
    \item \textbf{Enhanced observability and explainability:} Integration of interactive visualization and monitoring tools to support stakeholders in understanding and interpreting KG log analyses.
\end{itemize}

\paragraph{Future Directions and Research Opportunities.}  
Based on our survey, several promising avenues emerge:

\begin{itemize}
    \item \textbf{Standardization of KG query-log formats and meta-models:} To facilitate benchmarking, reproducibility, and tool interoperability.
    \item \textbf{Trust-aware analytics frameworks:} Development of metrics, propagation models, and query languages that explicitly account for reliability and provenance.
    \item \textbf{Automated pipeline orchestration:} Leveraging AI and workflow engines to automate preparation, curation, and storage, reducing manual effort and errors.
    \item \textbf{Cross-domain integration:} Combining KG query logs with other forms of user-generated content and web logs to enable richer insights and predictive analytics.
    \item \textbf{Explainable analytics:} Embedding interpretable models, visualization dashboards, and traceability mechanisms to ensure transparency and accountability.
\end{itemize}
 
Our survey demonstrates a clear lack of end-to-end, trust-aware solutions for KG query-log management. This motivates our proposed architecture, which addresses the full lifecycle of KG logs—from preparation and curation to storage, trust modeling, and analytics. By integrating trust, observability, and structured analytics, such a solution not only fills existing gaps but also sets a foundation for systematic, replicable, and explainable KG query-log exploitation. The detailed design of this architecture is presented in the next section.

\section{Conclusion}
Knowledge Graph (KG) query logs represent a critical but still underexplored resource for understanding and improving KG usage. Our survey has highlighted that, unlike relational database or web search logs, KG query logs have been primarily studied from a descriptive perspective, focusing on the structure, frequency, and graph-specific characteristics of SPARQL queries. Only a few works have ventured beyond analysis to propose applications such as query recommendation, reformulation, or benchmarking.

Despite these efforts, significant gaps remain. Raw query logs are largely impractical for direct exploitation, as they require structured pipelines that encompass preparation, curation, storage, and usage. Trust, which is a central concern in open-world and collaborative environments, has been almost completely overlooked in existing studies, even though it directly influences the reliability of analytics and downstream applications. Moreover, the lack of standardized tools, ontologies, and visualization frameworks limits the reproducibility, observability, and explainability of KG log analytics.

To address these gaps, our work emphasizes the need for a comprehensive, usage-agnostic, end-to-end architecture for KG query-log analytics. Such an architecture must support all stages of the log lifecycle, from metadata extraction and cleaning to curation, storage, and analytics. Integrating trust as a first-class concern allows the annotation of SPARQL queries and KG logs with trust values, providing a robust foundation for reliable usage. Coupled with visualization tools and modeling frameworks, this approach enhances transparency, observability, and explainability, supporting both research and industrial applications.

Overall, our findings underline that KG query logs are a valuable emerging element of user-generated content in knowledge systems, yet their potential remains untapped. By providing a systematic framework that incorporates trust, pipeline operations, and analytics support, this work lays the foundation for future research and practical applications. Future studies may focus on standardizing log formats, developing dedicated meta-models and ontologies, and designing advanced analytics and visualization tools to fully exploit the richness of KG query logs.

\section{Acknowledgments}
We extend our gratitude to Professor BELLATRECHE Ladjel and Dr. KHOURI Selma for their invaluable guidance, insightful ideas, and contributions.

\bibliographystyle{unsrtnat}
\bibliography{references}  

\end{document}